\documentclass[aps,prb,reprint,twocolumn,amsmath,amssymb,superscriptaddress,citeautoscript]{revtex4-1}
\usepackage{graphicx}
\usepackage{color}
\renewcommand{\vec}[1]{\mathbf{#1}}

\begin{document}
\title{Equilibrium ultrastable glasses produced by random pinning}

\date{\today}

\author{Glen M. Hocky}
\affiliation{Department of Chemistry, Columbia University, 
3000 Broadway, New York, New York 10027, USA}

\author{Ludovic Berthier}
\affiliation{Laboratoire Charles Coulomb, UMR 5221, 
CNRS and Universit{\'e} Montpellier 2, Montpellier, France}

\author{David R. Reichman}
\affiliation{Department of Chemistry, Columbia University, 
3000 Broadway, New York, New York 10027, USA}

\begin{abstract}
Ultrastable glasses have risen to prominence due to their potentially 
useful material properties and the tantalizing possibility of a general 
method of preparation via vapor deposition. Despite the importance of this
novel class of amorphous materials, numerical 
studies have been scarce because achieving ultrastability in atomistic 
simulations is an enormous challenge. Here we bypass 
this difficulty and establish that randomly pinning
the position of a small fraction of particles inside an equilibrated
supercooled liquid generates ultrastable configurations at essentially 
no numerical cost, while avoiding undesired structural changes due 
to the preparation protocol. Building on the analogy with 
vapor-deposited ultrastable glasses, we study the melting kinetics of these 
configurations following a sudden temperature jump into the liquid phase. 
In homogeneous geometries, we find that enhanced kinetic stability is 
accompanied by large scale dynamic heterogeneity, while a competition 
between homogeneous and heterogeneous melting is observed when a liquid 
boundary invades the glass at constant velocity. Our work demonstrates 
the feasibility of large-scale, atomistically resolved, and experimentally 
relevant simulations of the kinetics of ultrastable glasses.
\end{abstract}
\maketitle

\section{Introduction}

Recently, glasses with remarkable thermodynamic and kinetic stability have 
been prepared by vapor deposition on substrates maintained below the 
conventional glass transition temperature, $T_g$, of the bulk 
liquid~\cite{Swallen-Science2007,Dalal-JCPL2012,Guo-NatMat2012,newmark2014}.
It is estimated that these ``ultrastable'' glasses occupy states 
that are so low in the energy landscape that it would take several decades 
of conventional annealing of amorphous samples to prepare 
materials with equivalent properties~\cite{Swallen-Science2007}.
In addition to potential technological 
applications~\cite{Swallen-Science2007,Ediger-JCP2012}, 
these novel materials raise new challenges for theory and thus opportunities 
for gaining a deeper theoretical understanding of amorphous 
materials~\cite{Berthier-RMP2011}.

To rationalize the formation of ultrastable glasses,
Ediger and coworkers have hypothesized that deposition 
on a cold substrate combined with enhanced mobility at the free surface 
allows the system to effectively burrow into deeper free energy 
minima~\cite{Swallen-Science2007}. Computational studies using 
facilitated lattice models~\cite{Leonard-JCP2010}, 
simulations of Lennard-Jones systems~\cite{Singh-NatMat2013,Lyubimov-JCP2013}, 
as well as theoretical analysis based on random first order transition
theory~\cite{Stevenson-JCP2008,Wolynes-PNAS2009}
have demonstrated phenomenology in harmony with these ideas. 
By inhabiting deeper basins on the energy landscape, stable 
glasses may be closer to the putative Kauzmann 
or ``ideal glass transition'' temperature, $T_K$. 
This idea has been quantified by experimental estimation of fictive 
temperatures lying well below $T_g$~\cite{Swallen-Science2007}. 
Ultrastable glasses may thus provide an experimental means of producing an 
amorphous material with very low configurational entropy.

As difficult as it is to experimentally produce annealed glasses at 
temperatures that approach $T_K$, it is all the more so {\em in silico}. 
Even the most powerful computers and computational methodologies cannot 
simulate model supercooled liquids that approach $T_g$, let alone $T_K$, 
although work mimicking vapor deposition 
protocols~\cite{Singh-NatMat2013,Lyubimov-JCP2013} as well as the use of 
biased sampling of trajectories~\cite{Jack-PRL2011,Speck-PRL2012}, 
have moved us closer to this goal.
Recently, theoretical investigations have been put forward that potentially 
make accessing and testing the behavior of ``ideal glass'' states possible
using concepts borrowed from studies of fluids in porous media. 
Specifically, the physics of a fluid in the presence of 
a small fraction of randomly pinned particles has been 
shown numerically~\cite{Kob-PRL2013} and 
theoretically~\cite{cammarota-PNAS2012} to share essentially the same 
glassy physics as bulk supercooled liquids, which justifies more generally
the use of this particular strategy in the context of general 
studies of glass formation~\cite{Kob-PRL2013,cammarota-PNAS2012,Kim2003,Jack-PRE2012,karmakar-pinned,Tarjus2012,chiara1,chiara2,Berthier-static-PRE2012,jack2013,Chakrabarty-arxiv2014,Fullerton-arxiv2013,Gokhale-arxiv2014}. 

Random pinning presents two distinctive features 
with respect to bulk liquids that are central to 
our study. First, glassy dynamics and the transition to
glassy states occur at temperatures that are higher than in 
bulk~\cite{Kim2003}, because pinning a fraction $f$ of the particles restricts 
the available configurational space~\cite{Kob-PRL2013,cammarota-PNAS2012}. 
Second, configurations produced by randomly pinning 
particles within a thermalized supercooled liquids are, 
by construction, at thermal equilibrium~\cite{krako}. 
Together, these two features suggest 
that equilibrium configurations created by random pinning 
{\it correspond to a degree of supercooling 
at finite $f$ that cannot be achieved by conventional means.} 
Our central working hypothesis is that these pinned systems must 
share many properties with ultrastable glasses formed in the laboratory 
via nonequilibrium vapor deposition techniques. We shall demonstrate 
explicitely the validity of this hypothesis. 

While the kinetic stablity of pinned configurations does not shed 
light on the specific properties of the vapor deposition
process, it opens up the possibility to perform detailed 
microscopic analysis of the melting process observed when an 
ultrastable glass configuration is heated and melts back to the equilibrium
liquid. This represents the main goal of the present work.  

Crucially, because we obtain pinned configurations directly from bulk fluids, 
preparation of stable configurations is numerically very easy. 
This allows us to explore large systems sizes, long relaxation timescales, 
and various geometries, while automatically avoiding any undesired structural 
changes due to preparation protocol such as anisotropy or concentration 
fluctuations~\cite{Dawson-JCP2012,Lyubimov-JCP2013,Singh-JCP2013}. 
Mimicking experimental work characterizing ultrastable glasses, 
we first demonstrate ultrastability using conventional calorimetry. We then 
explore with atomistic resolution the kinetics of melting of stable 
glasses into the liquid phase following a sudden temperature jump, both 
in homogeneous and inhomogeneous geometries. 

\section{Model and preparation of ``equilibrium
pinned glasses''}

We study the properties of a two-dimensional binary Lennard-Jones 
mixture, employing a 65:35 mixture of particles with 
interactions parameters as in the model of Kob-Andersen~\cite{Kob-PRL1994},
which has been previously shown to be stable against 
crystallization and serves as a model supercooled 
liquid~\cite{Bruning-JPCM2009,Hocky-JCP2013}. Studying the two-dimensional 
system allows us to more directly visualize the spatial 
variation in observables. Exploratory studies of the three-dimensional
system suggest that our results for calorimetric measurements and 
kinetic stability do not sensitively depend on dimensionality. 
Bidimensional systems offer the additional advantage that 
larger system sizes can be studied, which proves a decisive advantage 
when studying spatial kinetic heterogeneities and inhomogeneous geometries. 

The Kob-Andersen Lennard-Jones system~\cite{Kob-PRL1994} 
is a binary system of 
particles with pairwise Lennard-Jones interactions, such that particles 
$i$ and $j$ separated by a distance $r_{ij}$ interact with the potential 
$V(r_{ij})=\epsilon_{\alpha\beta}\left ( (\sigma_{\alpha\beta}/r_{ij})^{12}-(
\sigma_{\alpha\beta}/r_{ij})^6 \right )$, where $\alpha$ and $\beta$ 
represent the particle type, $A$ or $B$ for particle $i$ and $j$ 
respectively. Interaction parameters are given by $\epsilon_{AA}=1$, 
$\epsilon_{AB}=\epsilon_{BA}=1.5$, $\epsilon_{BB}=0.5$, and $\sigma_{AA}=1$, 
$\sigma_{AB}=\sigma_{BA}=0.8$, $\sigma_{BB}=0.88$. The 
interaction between particle types $\alpha$ and $\beta$ is truncated and 
shifted up at distance 2.5$\sigma_{\alpha\beta}$. All particles have equal 
mass $m$. Energies, distances, and times are reported in reduced units 
proportional to $\epsilon_{AA}$, $\sigma_{AA}$ and $\tau=\sqrt{m 
\sigma_{AA}^2/\epsilon_{AA}}$. As stated above, we study the 
two-dimensional variant of the model, 
for which a 65:35 ratio of $A$ to $B$ particles was previously 
shown to be a robust model supercooled liquid, resistant to 
crystallization~\cite{Bruning-JPCM2009,Hocky-JCP2013}. 

We generated configurations using 
Molecular Dynamics at a series of decreasing temperatures following the 
procedure of Ref.~\onlinecite{Hocky-JCP2013} to generate configurations 
at density $\rho=1.2$ with $N=10000$ down to $T=0.45$, and then generated 
configurations at $T=0.425$ by running simulations on these configurations 
for an additional $1.25\times10^{5}\tau$ using an integration time-step of 
${\rm d}t=0.005\tau$. In all cases, simulations were performed using LAMMPS
 \cite{Plimpton-JCP1995}, and the temperature was maintained using a 
Nos\'{e}-Hoover thermostat with a time constant of $100{\rm d}t$ 
\cite{Martyna-JCP1992}.
We prepared many independent 
equilibrated supercooled configurations with $N=10^4$ 
and density $\rho=1.2$ at initial temperature $T_i=0.425$. For the bulk,  
the mode-coupling temperature (which roughly 
coincides with the computer glass transition) is
$T \approx 0.4$. 

In these equilibrated configurations, we 
fix the position of a percentage $f = 100 \tilde{f}$ of particles. 
This produces equilibrium configurations at state point $(f,T_i)$, which are 
our numerical analog of the samples that are vapor-deposited
below the glass temperature, in the sense that both protocols
produce amorphous configurations with lower energy
at the preparation temperature than configurations prepared 
by slowly re-annealing these samples.
We wished to perform simulations with a fraction $\tilde{f}$ of 
particles fixed in place in a ``uniformly random'' manner similar in spirit 
to Ref.~\onlinecite{Kob-PRL2013} to avoid sampling problems arising from 
localized extra-slow dynamics due to clusters of pinned particles. In order 
to do this, we chose a minimum distance between pinned particles 
$d_{min}(\tilde{f})$ and then randomly picked a set of particles commensurate 
with this minimum pair difference. In practice, we were able to do this 
for the $\tilde{f}$ of interest by using $d_{\rm min}(\tilde{f})=0.85 
d_{\rm avg}(\phi={0.55})$, where $d_{\rm avg}(\phi)$ is the diameter $\tilde{f} 
N$ particles would have if they were placed in the same box with packing 
fraction $\phi=\tilde{f} \rho \pi (d_{\rm avg}/2)^2 $. In this way the 
radial distribution function for the pinned particles is similar to that 
of a simple liquid with packing fraction $\phi=0.55$, which was chosen 
heuristically. The results presented in this work appear robust to the 
specific choice here, or indeed if the particles to pin are chosen 
totally at random,  and this particular sampling should be viewed as
a numerical convenience with no incidence on our results. 

To characterize these configurations, we either use finite rates 
to heat/cool them and study their calorimetric properties, 
or sudden temperature jumps into the liquid phase, $T > T_i$,
to follow the kinetics of ``melting'' of the glass into the liquid. 
We refer to our pinned samples as ``pinned glasses,'' even
though they actually correspond to equilibrium samples.
These deeply supercooled states come at no numerical cost.  
Indeed, the ordinary challenge of finding a good configuration 
satisfying a difficult constraint (such as low temperature) is reversed 
here because we define the hard constraint {\em after} the standard bulk 
configuration has been prepared (similar to the idea of {\em planting}, 
see e.g. Ref.~\onlinecite{Krzakala-PRL2009}). 

In order to perform the heating and cooling experiments shown below in
Fig.~\ref{fig:annealing}, twelve simulations were run for every value 
of $f$ and $\gamma$, by choosing four uncorrelated 
sets of $\tilde{f} N$ random particles 
from three independent starting configurations. These 
configurations were heated from $T_i=0.425$ to $T=2.0$ with the thermostat 
temperature raised in a linear fashion at a rate $\gamma$. The samples were 
subsequently cooled back to $T=0.425$ and reheated to $T=2.0$, also at a 
rate $\gamma = dT / dt$. 

For other data in this work, the thermostat temperature was 
instantly raised to a ``melting temperature'' $T$ into the 
liquid phase. For any data point or 
curve shown here, at least 8 simulations were performed, with 4-6 
independent configurations using 2-5 sets of randomly pinned particles. 
The simulation temperature stabilized at $T$ by 1000${\rm d}t$, corresponding 
to about 10 times the thermostat relaxation time. For all of these simulations, 
the simulation integration time-step was kept at ${\rm d}t=0.005$ 
without any adverse consequences.

\section{Calorimetric measurements establish 
ultrastability of randomly pinned samples}

\begin{figure}
\centering
\includegraphics[scale=0.55]{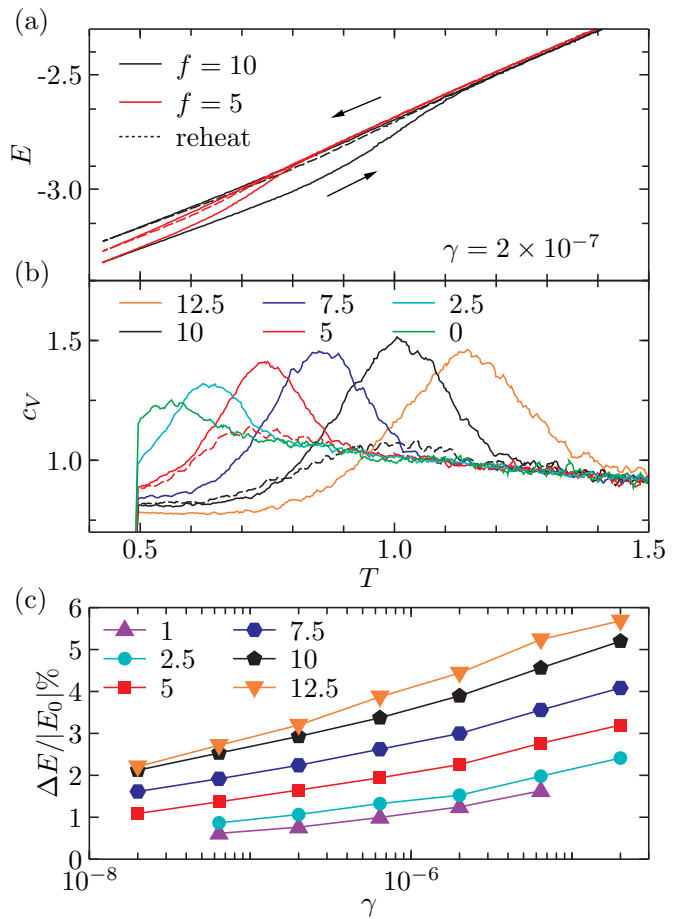}
\caption{{\bf Ultrastability
of pinned configurations via calorimetric measurements.}
(a) Average energy per particle of samples created with $f=5$ 
or $f=10$ heated from $T_i=0.425$ to $T=2.0$ and then 
cooled back down at a constant rate $\gamma$. The conventional 
glass preparation by cooling and the directly pinned configurations
reside on different glass branches, the difference increasing with $f$. 
(b) Heat capacity per particle measured upon 
heating the ultrastable configurations (solid lines)
displays a much higher peak as the glass transition is crossed
than the one of conventional glasses
(dotted lines, for $f=5$ and $f=10$).
(c) Energy gap at $T = T_i$ between directly pinned 
configurations and after cooling from the liquid at rate $\gamma$ 
for different pinning fractions $f$. It would take 
orders of magnitude slower cooling rates to prepare conventional 
glasses equivalent to pinned samples.}
\label{fig:annealing}
\end{figure}

To quantify the thermodynamic stability of our pinned samples, 
we mimic the experimental technique of scanning calorimetry.
To this end we heat, then cool, then reheat the 
sample again at constant rate $\gamma=dT/dt$ from initial temperature $T_i$ to 
a maximum temperature of $T=2.0$. In Fig.~\ref{fig:annealing}(a) we show 
the results of this temperature scans for samples with $f=5$ and $f=10$. 
Upon initial heating, the system stays on the 
``stable glass branch'' until the glass transition temperature is crossed. 
The system then melts and returns to the equilibrium 
liquid branch. On 
cooling, the samples do not return to the stable glass branch but to 
a different one which is much higher in energy. 
Little hysteresis is observed upon reheating, showing 
that the glass configurations produced by conventional
cooling are much less glassy. We also observe that the 
simulations for larger $f$ need to be heated to a much higher temperature
 before they ``melt'', 
confirming that the glass temperature increases with $f$. 
The difference in energy between the original sample 
and re-annealed samples also increases with $f$. Differential scanning 
calorimetry experiments on ultrastable glasses 
look quite similar, in the sense that the experimental conditions 
under which these glasses were prepared determine how 
different the two glass branches are~\cite{Swallen-Science2007}. 
The data in Fig.~\ref{fig:annealing} are also direct 
proof that randomly pinned glasses lie very deep in the 
energy landscape, in the sense that their energy is much lower than
the one of slowly re-annealed configurations. 

By taking the numerical derivative of the energy, we extract 
the heat capacity $c_V$ 
for pinned samples, shown in Fig.~\ref{fig:annealing}(b). 
We see that the peak temperature signaling the glass transition 
moves to higher temperature with increasing $f$, and the peak 
height also grows with $f$. By contrast,
heat capacities for the re-annealed samples (dotted) display 
a much smaller peak, providing verification that 
ordinary annealed samples occupy much higher energy states. 
The peaks in $c_V$ of the re-annealed glasses 
lie approximately under those from the initial heating. This
is in contrast to what is seen for ultrastable glasses, 
where the re-annealed samples melt at lower temperature 
than the vapor deposited ones~\cite{Swallen-Science2007}. 
This is probably because 
our numerical heating rates are much faster than in experiments, 
resulting in $c_V$ peaks that are very broad, 
and thus much harder to resolve. 

The energy gap between stable and re-annealed pinned glasses 
depends on the annealing rate, and it shrinks for 
slower cooling (by definition the gap should vanish when $\gamma \to 0$). 
In Fig.~\ref{fig:annealing}(c) we show the energy gap $\Delta E$ measured 
at $T=T_i$, normalized by the average initial energy
$E_0$. In all cases, we observe $\Delta E > 0$,  with gaps that increase 
with $f$. Extrapolation of these curves, even for 
modest $f$, would suggest that only cooling rates smaller by 
several orders of magnitude could produce similarly stable 
configurations. This directly demonstrates the advantage of 
using pinned glasses, which can truly be described 
as extremely ``old'' glasses that we obtain at essentially no 
numerical cost, in excellent analogy with vapor-deposited 
glasses~\cite{Swallen-Science2007}. 

The results presented in this section demonstrate the ultrastability of 
pinned glasses. We recall that by construction these 
configurations represent equilibrium samples at the state 
point ($f,T_i$), whose structural features are thus
not particularly illuminating. Therefore our approach provides no 
useful information about the structure of ultrastable glasses 
in general. Remarkably, though, it allows us to study 
in great detail the non-equilibrium relaxation processes following a sudden
heating of these glasses into the liquid phase, a phenomenon of obvious
experimental relevance.  

\section{Kinetics of homogeneous melting of 
ultrastable pinned glasses into the liquid}

We now study the response of pinned glasses 
after a sudden change in temperature to $T > T_i$, 
the subsequent dynamics occurring fully out of equilibrium. 
We compute the self-overlap function for the 
$N_u$ unpinned particles, $q_s(t,t_w) = 
\frac{1}{N_u} \langle \sum_{m} q_m(t,t_w) \rangle$ with 
\begin{equation}
\label{eq:overlap_i}
q_m(t,t_w) = \theta(|\vec{r}_m(t+t_w) - \vec{r}_m(t_w)|-a),
\end{equation}
where $t_w$ is the waiting time since the temperature jump, 
$\theta(x)$ is the step function which is unity 
for $x \leq 0$ and zero for $x>0$, and ${\bf r}_m(t)$
denotes the position of particle $m$ at time $t$.
We choose $a=0.22$ such as the self-overlap relaxation time 
$\tau_s$, defined as $q_s(\tau_s,0) \equiv 1/e$, matches the relaxation 
time $\tau_\alpha$ in equilibrium conditions and as defined 
in a previous work~\cite{Hocky-JCP2013}. 

\begin{figure}
\centering
\includegraphics[scale=0.95]{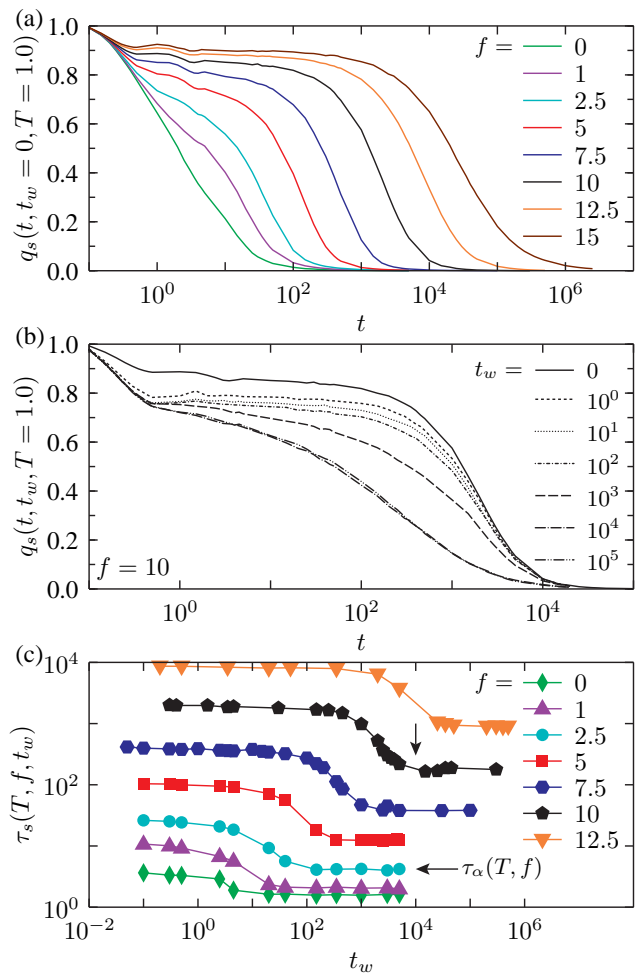}
\caption{{\bf Kinetics of homogeneous melting after temperature jump 
to liquid phase.}
(a) Self-overlap measured immediately ($t_w=0$)
after instantaneous heating to $T=1$ for different $f$, with $T_i=0.425$. 
Correlation functions decay more slowly as $f$ increases, with an 
increasing plateau height.
(b) Waiting time dependence of the self-overlap 
for $f=10$ after sudden jump to $T=1$. Relaxation 
accelerates as the glass melts into the liquid, and converges towards
equilibrium relaxation.
(c) Waiting time dependence of the overlap relaxation times $\tau_s$ 
for various $f$ after heating to 
$T=1.0$. Examples of transformation time, $t_{\rm trans}$, and 
equilibrium relaxation time, $\tau_\alpha$, are 
indicated by vertical and horizontal arrows, respectively. }
\label{fig:melting}
\end{figure}

The relaxation of samples heated to $T$ becomes drastically slower as $f$ 
is increased, as shown for $T=1$ in Fig.~\ref{fig:melting}(a). In 
addition, a plateau develops and grows in height suggesting a more stable 
glass~\cite{Singh-JCP2013,Singh-NatMat2013}. 
When the waiting time increases, the particle dynamics accelerates
as the glass progressively melts into the liquid, 
see Fig.~\ref{fig:melting}(b). The 
system eventually relaxes towards the thermalized liquid
state where dynamics becomes stationary.
A comparison between the $t_w=0$ (glass melting) and 
$t_w \to \infty$ liquid relaxation data shows that (i) the 
relaxation time of the liquid is faster, (ii) its time correlation function
shows more pronounced deviation from exponential decay, 
(iii) and the intermediate plateau height is much lower.
While (i) and (iii) reveal kinetic stability of the pinned glasses, 
(ii) is more surprising at it could naively suggest that the melting
relaxation process is less heterogeneous than the equilibrium 
dynamics. We shall see below that the opposite is actually true
in the sense that the nonequilibrium melting process is characterized
by a dynamic correlation lengthscale which is larger than the 
equilibrium relaxation dynamics.

The overlap relaxation time $\tau_s(T,f,t_w)$ can serve as an indicator of 
the progress of glass melting. 
In Fig.~\ref{fig:melting}(c), we show the overlap 
relaxation time for each $f$ after waiting time $t_w$. The relaxation time 
is initially large, decays to the equilibrium value, and
eventually becomes independent of $t_w$.
We wish to define a transformation time $t_{\rm trans}$ to quantify the 
time where $\tau_s(T,f,t_w)$ has fully decayed to this plateau value,
in analogy to its experimental determination~\cite{Dawson-JCP2012,newmark2014}. 
However, the correct functional form of the $t_w$-dependence of 
$\tau_s(T,f,t_w)$ is unclear. Instead, we have observed 
that the alternative definition of the transformation time 
as $q_s(t_{\rm trans},0,T) \equiv 0.01$ 
serves as a consistent proxy for $t_{\rm trans}$
across all $f$ and $T$ and is a simpler and more accurate tool for 
the extraction of $t_{\rm trans}$ than is inspection or fitting 
of the $\tau_s$ curves, and so we adopt this practical 
definition of $t_{\rm trans}$. 

With increasing $f$, we observe a substantial increase in $t_{\rm trans}$ 
relative to that for unpinned samples. For example, the data in 
Fig.~\ref{fig:melting}(c) show an increase of a factor of $\approx 10^3$.
However, the equilibrium relaxation time $\tau_\alpha$ also increases
with $f$. It is sensible to define a ``stability ratio'', 
${\cal S} = t_{\rm trans}/\tau_\alpha$, to characterize the melting of 
ultrastable glasses, as done in a recent set of experiments~\cite{newmark2014}.

\begin{figure}
\centering
\includegraphics[scale=0.9]{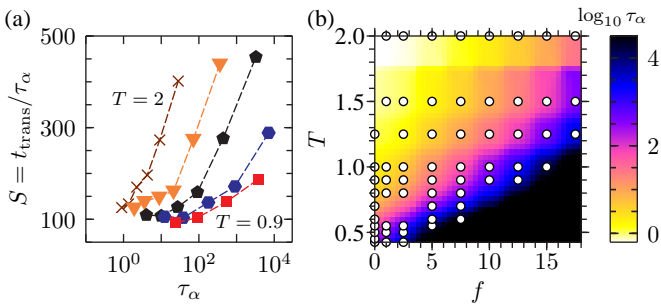}
\caption{{\bf Evolution of transformation and equilibrium relaxation times.}
(a) Evolution of stability ratio ${\cal S} = t_{\rm trans}/\tau_\alpha$ with 
equilibrium relaxation time $\tau_\alpha$ as $f$ is varied, for 
temperatures $T=\{2.0, 1.5, 1.25, 1.0,0.9\}$ (from left to right) 
with $f=\{5,7.5,10,12.5,15,17.5\}$ where accessible, with $T_i=0.425$. 
The stability ratio increases with the glass thermodynamic stability.
(b) Contour plots of the equilibrium 
relaxation time $\tau_\alpha$ in the $(f,T)$ phase diagram.
Symbols indicate the location of the simulations performed
to obtain the data shown in (a). Pinned glasses produced at 
$T_i=0.425$ have equilibrium relaxation times that lie in a (black) region that is not accessible at equilibrium by conventional means.}
\label{fig:taus}
\end{figure}

In Fig.~\ref{fig:taus}(a) we present the evolution of 
the stability ratio for all $(f,T)$ points studied. 
The corresponding wide range of $\tau_\alpha$ is 
highlighted in Fig.~\ref{fig:taus}(b).
We find that ${\cal S}$ is always larger than about 100, 
with a maximum value reaching ${\cal S} \approx 500$ for the most stable 
systems. For given preparation and melting
temperatures $(T_i,T)$, we find that ${\cal S}$ increases
with $f$. This result is expected since the 
initial state then corresponds to a pinned glass that is 
increasingly stable, as demonstrated in 
Fig.~\ref{fig:annealing}. In other words, the kinetic stability ratio
${\cal S}$ measured in nonequilibrium melting protocols 
is strongly correlated with the thermodynamic ultrastability 
revealed by calorimetric measurements. 

Our data also indicate that 
increasing the melting temperature $T$ decreases both $\tau_\alpha$
and $t_{\rm trans}$, but their ratio remains approximately constant. 
This suggests that ${\cal S}$ is a robust way to compare 
the stability of samples with different preparations in our simulations. 
This conclusion is not obvious, as experiments report that 
the stability ratio in fact increases with the melting temperature 
$T$ for a given glass preparation. A plausible explanation 
is that although our pinned glasses correspond to equilibrium, low 
energy states, we melt them in a relatively high temperature regime 
such that both $\tau_\alpha$ and $t_{\rm trans}$ can be numerically 
measured. Instead, glass melting in experiments is usually 
performed at much lower temperatures close to the experimental $T_g$.

While ${\cal S}$ is substantially larger for the 
equilibrium pinned glasses than for conventionally prepared 
(through slow annealing) glass
configurations (our tests indicate a gain of about 10), we do
not find stability ratios ${\cal S}$ in the range $10^3 - 10^6$ reported 
experimentally~\cite{newmark2014}. Our results reveal in fact 
that the stability ratio increases only weakly with increasing 
the glass stability, since the equivalent of several decades of 
annealing increases ${\cal S}$ by less than a decade. We can 
only speculate that extrapolating the data shown  
in Fig.~\ref{fig:taus}(a) to pinned glasses obtained from 
bulk samples prepared at even lower temperatures  
could plausibly yield stability ratios comparable to the 
ones determined experimentally. 

Finally, note that our equilibrium pinned 
samples do not display any structural change with respect to the bulk,
such as anisotropy, layering, or compositional fluctuations
which might occur during vapor 
deposition~\cite{Dawson-JCP2012,Lyubimov-JCP2013,Singh-JCP2013},
although they do possess non-trivial many-body 
correlations~\cite{Fullerton-arxiv2013}.
Such structural changes could also contribute 
to increasing the stability ratio measured experimentally. \\

\section{Enhanced kinetic stability is accompanied by 
large-scale  dynamic heterogeneity}

\begin{figure}
\centering
\includegraphics[scale=0.95]{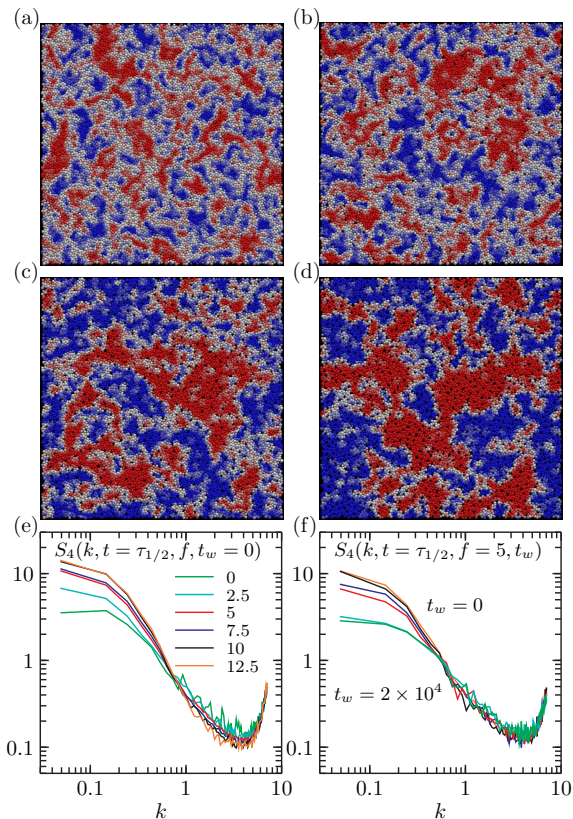}
\caption{{\bf Large-scale dynamic heterogeneity during 
melting of pinned glass to pinned liquid.}
(a-d) Representative snapshots of melting of pinned glasses 
at $T=0.9$, with $f=0$, 2.5, 7.5, and 12.5 from (a) to (d).
Each particle is colored by the local value of the self-overlap 
(averaged in a disc of radius 1.5), 
ranging from blue for $q_m=1$ (glass) to red with $q_m=0$ (liquid).
Times are chosen so that the average overlap is about 0.5, 
pinned particles are not shown. 
(e) The corresponding four-point dynamic structure factor 
increases and reveals longer ranged spatial correlations with 
increasing $f$. 
(f) Waiting time dependence of four-point dynamic structure 
factor decreases showing that dynamic heterogeneity 
are reduced as the glass melts into the liquid.}
\label{fig:spatial}
\end{figure}

Atomistic simulations offer the possibility to observe how the 
melting process of the glass occurs in homogeneous geometries.
In Fig.~\ref{fig:spatial} we show
the spatial distribution of $q_m(t=\tau_{1/2},t_w=0,T)$, where 
$\tau_{1/2}$ is the time at which $q_s(t=\tau_{1/2},t_w=0)=1/2$, and
the color of the particles represents $q_m$ coarse-grained 
over a local region of radius 1.5, for clarity purposes. 
Fig.~\ref{fig:spatial}(a-d) show snapshots for $f=$ 0, 2.5, 7.5, 
and 12.5, which allow us to observe how the stable glass initial configuration
(in blue) progressively melts into the liquid (in red). 
We observe that liquid pockets emerge at some random location 
in the glass, and grow in size until the whole system 
has melted. Clearly as $f$ (and thus the transformation time) 
increases, we see that the ``nucleation'' regions where the 
liquid first appears become sparser, so that the 
size of the liquid/glass domains observed  
at time $t=\tau_{1/2}$ grow in size. Using the language of equilibrium 
glass transition studies~\cite{Berthier-RMP2011}, the melting dynamics 
becomes spatially correlated over larger distances when 
kinetic stability increases. 

To quantitatively validate this visual impression, 
we compute the four-point dynamical structure factor which measures 
the scattering off of these dynamic domains,
\begin{equation}
S_4(k,t,t_w) = \frac{1}{N_u}  \langle \sum_{m,n} q_m q_n  e^{i 
\vec{k}\cdot (\vec{r}_m(0)-\vec{r}_n(0))}  \rangle,
\end{equation}
where $q_m$ is shorthand for $q_m(t,t_w)$ defined in 
Eq.~(\ref{eq:overlap_i}). The resulting profiles shown for one temperature 
in Fig.~\ref{fig:spatial}(e) have larger spatial variation measured 
by $S(k\rightarrow 0,t)$ with increasing $f$. This reveals a growing 
non-equilibrium length scale for dynamically heterogeneous melting. If 
we increase the waiting time, we observe from 
Fig.~\ref{fig:spatial}(f) that the dynamic heterogeneity 
decreases as the sample transforms back into the liquid. Finally, for 
a given value of $f$, we find that $S_4$ also grows as $T$ is 
decreased (data not shown). The behavior of the four-point susceptibility 
$\chi_4(t,t_w)$ can be directly 
deduced from the $k \to 0$ behavior of $S_4(k,t,t_w)$, 
and its evolution simply mirrors the one described above 
for $S_4(q,t,t_w)$.

Figure~\ref{fig:spatial} shows that glass melting
is spatially heterogeneous, with a 
correlation length scale increasing with the glass stability. 
Additionally, dynamics is correlated over larger distances (and 
is thus spatially more heterogeneous) for $t_w=0$ (where 
time correlations are nearly exponential) than for equilibrium 
(where time correlations are stretched). Another interesting point
is the growth of spatial correlations with $f$ at constant
$T$ observed in Fig.~\ref{fig:spatial}, which contrasts with 
recent numerical studies suggesting a much weaker 
effect on the dynamic correlation length in equilibrium 
conditions~\cite{jack2013}. 
Overall, our data thus suggest that the nonequilibrium 
melting process of the glass into the liquid differs qualitatively
from the ordinary relaxation dynamics observed in thermal 
equilibrium conditions. 

Physically, our results appear consistent with both a 
dynamic picture where relaxation is first triggered 
by a sparse population of ``defects'' and then 
propagates in space via dynamic 
facilitation~\cite{Jack-PRL2011,Garrahanreview}, and with a
thermodynamic picture where the melting occurs 
via ``nucleation and growth'' of the liquid into the glass,
as envisioned by random first order transition 
theory~\cite{Bouchaud-JCP2004,Wolynes-PNAS2009,Xia2000,Krzakala-JCP2011-1,Krzakala-JCP2011-2}. Using the latter approach, quantitative phase diagrams 
have been obtained for the situation we study. This analysis suggests the
existence of a first-order transition separating the randomly pinned 
configuration from the equilibrium liquid~\cite{chiara2}. In this
view, our results should represent a genuine ``melting'' process from 
one phase to another, thus justifying our frequent use of the
nucleation theory language throughout the article. 
Although this view would naturally account for the qualitative 
features reported in the present study, our work
should not be taken as a quantitative indication 
of the {\it existence} of the thermodynamic 
phase transitions as determined in the theoretical 
analysis~\cite{chiara2}.

\section{Coexistence of homogeneous melting and
front propagation in inhomogeneous geometries} 

Experiments on ultrastable glasses in inhomogeneous geometries 
have demonstrated that when the stable glass is capped with 
a liquid interface, melting can be initiated from the 
liquid boundary which propagates at constant velocity into the 
glass~\cite{Sepulveda-JCP2012,Sepulveda-JCP2013}. 
These observations suggest that ``heterogeneous glass melting'' 
can compete with the ``homogeneous melting'' process studied above,  
in analogy with nucleation across a first-order phase transitions. 
This analogy can be rationalized by the theoretical analysis in 
Ref.~\onlinecite{chiara2}. 

\begin{figure}
\centering
\includegraphics[scale=0.95]{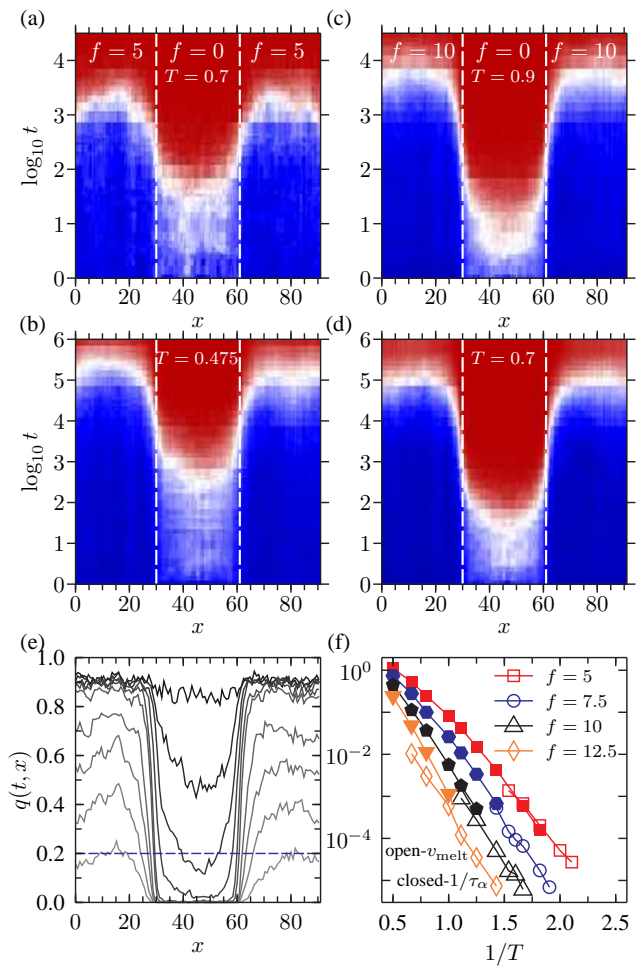}
\caption{{\bf From bulk to heterogeneous melting
via front propagation of the liquid phase.} 
(a-d) Relaxation profiles for systems where $f=5$ (a,b) or 
$f=10$ (c,d) everywhere except in the middle of the sample
(dashed lines) where $f=0$. The samples are heated to $T$ 
and the overlap profiles $q(t,x)$ are shown as a function 
of the time $t$ (color code as in Fig.~\ref{fig:spatial}).
Relaxation occurs rapidly in the center, and the liquid slowly invades the 
glass until homogeneous melting occurs and the interface disappears.    
(e) Time slices for the data in panel (d) for times increasing from 
top to bottom. 
(f) The temperature evolution of the melting front velocities 
$v_{\rm melt}$ essentially tracks that of 
the structural relaxation time, $v_{\rm melt} \sim \tau_\alpha^{-1}$.} 
\label{fig:interface}
\end{figure}

To investigate this situation, we produce a liquid/glass interface
by leaving a strip along the $y$ direction 
of the system with no pinning ($f=0$), 
while the rest of the system is pinned at finite $f$, as before. The unpinned 
region is the numerical analog of the liquid interface in experiments.
In this inhomogeneous geometry, we monitor 
the self-overlap profiles along the $x$ direction perpendicular 
to the liquid strip, $q(t,x) \equiv \langle q_m(t,t_w=0) \rangle_y$ 
for particles $m$ with horizontal positions in a strip of width 0.8 
centered at $x$. 
The results in Fig.~\ref{fig:interface}(a-d) 
demonstrate that relaxation occurs very rapidly in the unpinned region,
creating a liquid/glass (red/blue) interface, as desired.  
As time increases, the position of the interface (white)
moves and the melted section propagate 
into the pinned glass. When $t \sim t_{\rm trans}$, the pinned glass
may relax homogeneously and the interface 
disappears. The system is entirely fluid when $t \gg t_{\rm trans}$. 
Representative overlap 
profiles~\cite{Leonard-JCP2010,Sepulveda-JCP2013} are shown in 
Fig.~\ref{fig:interface}(e). We locate the ``propagating front''
from the overlap value $q(t,x_{\rm front})=0.2$, and  
fit its position for a given $(T,f)$ to a linear function 
of time, which defines the ``melting velocity'', 
$v_{\rm melt} = v_{\rm melt}(T,f)$.

The temperature dependence of $v_{\rm melt}$ is 
shown for four values of $f$ in 
Fig.~\ref{fig:interface}(f). At higher temperatures, it was not possible 
to observe a melting front over sufficient time range.
This is exemplified in Fig.~\ref{fig:interface}(a), 
where at $\log_{10}(t) \approx 3.2$, the sample 
has fully melted at a position far from the interface, indicating that 
the time scale for the bulk relaxation is shorter than the time taken 
for the front to propagate over a significant distance. 
Our numerical results indicate that 
$v_{\rm melt}$ is approximately proportional to $\tau_\alpha^{-1}$, see 
Fig.~\ref{fig:interface}(f), suggesting that the same relaxation mechanisms 
that govern melting front propagation allow for density relaxation in the 
pinned bulk. It is similarly found in experiments that $\ell \equiv 
v_{\rm melt} \tau_\alpha$ has a weak temperature dependence, and 
corresponds to a length scale of about $\approx 0.01 \sigma$, 
where $\sigma$ represents a molecular dimension~\cite{newmark2014,brazil}. 
In our simulations, $\ell$ has molecular dimensions, $\ell 
\approx 0.1-1 \sigma$, with essentially no temperature dependence. 
 
The observed competition between front propagation (speed $v_{\rm melt}$)
and homogeneous melting (timescale $t_{\rm trans}$) suggests a
maximum length scale over which heterogeneous melting can be 
observed, $\xi \approx v_{\rm melt} t_{\rm trans}$.
Interestingly this can be rewritten using the two quantities introduced 
above as $\xi = \ell \times {\cal S}$, showing that $\xi$ is 
mainly controlled by the absolute value of the stability 
ratio ${\cal S}$, because $\ell$ depends only weakly on temperature. 
Since the values observed in experiments can be up to
${\cal S} \approx 10^6$, a reasonable assumption of 
molecular dimensions for $\ell$ results  
in a maximal length scale for front propagation of
the order of microns, as found~\cite{newmark2014}.
Our prediction that $\xi \propto {\cal S}$ is 
experimentally verifiable. 

Interestingly, since this massive length scale $\xi$ 
directly results from kinetic stability, it is a priori not 
obvious that it could be interpreted as 
a large correlation length scale characterizing the melting 
of ultrastable glasses. Because we have observed an important
growth of the typical size of dynamic heterogeneity characterizing
the homogeneous glass melting in Fig.~\ref{fig:spatial}, it is however tempting 
to speculate that both observations are in fact related. If correct, 
this interpretation suggests that the typical size 
of dynamic heterogeneity observed during the melting of ultrastable
glasses in thick films can be as large as a micron, which would 
make its experimental observation much easier than  
equilibrium kinetic heterogeneities, whose typical size falls 
in the nanometer range. This suggests also that the melting of conventional
glasses is also characterized by dynamic length scales which are
potentially quite large and could then also be experimentally studied.    

\section{Discussion and perspectives}

We have shown that the random pinning of particles in a two-dimensional 
model supercooled liquid produces highly stable {\em in silico} glasses 
termed ``pinned glasses''. Our approach does not address 
specificities due to the experimental vapor deposition process, 
but allows instead a detailed exploration of the physical 
properties of ultrastable glasses, and we have studied more 
particularly the nonequilibrium process by which a stable 
glass melts back into the equilibrium liquid.  

We have demonstrated that pinned glasses behave in many ways as 
experimentally realized ultrastable glasses. 
Both systems lie much deeper in the energy landscape than ordinarily
prepared glass configurations, 
are characterized by peculiar calorimetric properties, and enhanced
kinetic stability. Both exhibit a competition
between homogeneous and heterogeneous glass melting, which
results, for inhomogeneous geometries, in a liquid front invading 
the glass over a large length scale controlled by the 
ratio ${\cal S} = t_{\rm trans}/\tau_\alpha$ 
between the bulk transformation time and the equilibrium relaxation time.  

An important difference between the two systems
is that pinned samples are formed in equilibrium, 
with no structural change induced by the preparation protocol, 
i.e. we can fully decouple stability and structure, 
in a way that might be experimentally realized in colloidal 
materials \cite{Curtis-Optics2002,Gokhale-arxiv2014}.  
We have attributed the larger stability ratio values 
measured experimentally both to this distinction
and to the fact that melting simulations are performed 
in a different temperature regime than in experiments.
Our work nevertheless shows that preparing 
energetically favorable configurations via random pinning is 
a promising route to understand the properties 
of vapor-deposited ultrastable glasses. Both types 
of systems thus appear as a type of amorphous materials
that cannot be prepared by conventional annealing protocols, which 
confers them with extraordinary physical properties. 

\acknowledgments

We thank M. Ediger, S. Dalal, and G. Biroli for stimulating 
conversations. LAMMPS simulations were executed and organized using the 
Swift parallel scripting language (NSF Grant No.~OCI-1148443) 
\cite{Wilde-Parallel2011} and were executed in part on resources 
provided by the University of Chicago Research Computing Center and 
on resources at the Texas Advanced Computing Center (TACC) provided 
through the Extreme Science and Engineering Discovery Environment 
(XSEDE), supported by NSF Grant No.~ACI-1053575. G.M.H was supported 
by NSF Grant No.~DGE-07-07425 and D.R.R. by CHE-1213247. 
The research leading to these results has received funding
from the European Research Council under the European Union's Seventh
Framework Programme (FP7/2007-2013) / ERC Grant agreement No 306845.

\bibliography{jcp3}

\end{document}